\documentclass[page-classic]{epl2} 

\graphicspath{{Images/}{figures/}}
\usepackage{textcomp}
\usepackage{latexsym}
\usepackage{amsmath}
\usepackage{amssymb}
\usepackage{color}

\definecolor{alizarin}{rgb}{0.82, 0.1, 0.26}
 \definecolor{BLACK}{gray}{0}
 \definecolor{WHITE}{gray}{1}
 \definecolor{RED}{rgb}{1,0,0}
 \definecolor{debianred}{rgb}{0.84, 0.04, 0.33}
 \definecolor{GREEN}{rgb}{0,0.7,0}
 \definecolor{green(ncs)}{rgb}{0.0, 0.62, 0.42}
\definecolor{armygreen}{rgb}{0.29, 0.33, 0.13}
 \definecolor{BLUE}{rgb}{0,0,1}
 \definecolor{CYAN}{cmyk}{1,0,0,0}
 \definecolor{MAGENTA}{cmyk}{0,1,0,0}
 \definecolor{YELLOW}{cmyk}{0,0,1,0}
 \definecolor{electricviolet}{rgb}{0.56, 0.0, 1.0}
 \definecolor{dukeblue}{rgb}{0.0, 0.0, 0.61}

\usepackage[normalem]{ ulem }
\usepackage{soul}

\title{Stripes instability of an oscillating non Brownian iso-dense suspension of spheres}
\shorttitle{Title} 

\author{Y. L. Roht\inst{1,2}\and I. Ippolito\inst{2} \and J.P. Hulin\inst{1}  \and D. Salin\inst{1} \and G. Gauthier\inst{1}}
\shortauthor{F. Author \etal}

\institute{
  \inst{1} Laboratoire FAST, Univ. Paris Sud, CNRS, Universit{\'e} Paris-Saclay, F-91405, Orsay, France\\
  \inst{2} Universidad de Buenos-Aires, Facultad de Ingenier\'{\i}a, Grupo de Medios Porosos, Paseo Col\'{o}n 850, 1063, Buenos
Aires, Argentina
}

\abstract{We analyze experimentally the behavior of a non-Brownian, iso-dense suspension of spheres submitted to periodic square wave oscillations of the flow in a Hele-Shaw cell of gap $H$. We do observe an instability of the initially homogeneous concentration in form of concentration variation stripes transverse to the flow. The wavelength of these regular spatial structures scales roughly as the gap of the cell and is independent of the particle concentration and of the period of oscillation. This instability requires large enough particle volume fractions $\phi\geq 0.25$ and a gap large enough compared to the sphere diameter ($H/d \geq 8$). Mapping the domain of existence of this instability in the space of the control parameters shows that it occurs only in a limited range of amplitudes of
the  fluid displacement. The analysis of the concentration distribution across the gap supports  a scenario of particle migration towards the wall followed by an instability due to a particle concentration gradient with a larger concentration at the walls. In order to account for the main features of this stripes instability, we use the theory of longitudinal instability due to normal stresses difference and recent observations of a dependence of the first normal stresses difference on the particle concentration.}

\begin{document}

\maketitle
\date{\today}

\section{Introduction}
\label{intro}
Since  pioneering work on viscous resuspension \cite{leighton87} and the observation of particles migration in flowing dense neutrally buoyant suspensions \cite{phillips92}, the so-called shear induced particle migration has been intensively studied both theoretically \cite{nott94} and experimentally \cite{butler99}. This migration was modeled using the so-called shear induced migration model \cite{leighton87,phillips92} or introducing normal stresses within the flowing suspension \cite{nott94}. It was soon recognized that normal stresses differences can lead to instabilities either longitudinal to the flow direction \cite{hinch92} or perpendicular to it \cite{brady02}. However, some properties of suspension flows  are not yet fully understood, such as the dynamics of the shear induced migration in axisymmetric Poiseuille flows at moderate volume fractions ($\phi \leq 0.3$) \cite{snook16}. Moreover, recent numerical simulations~\cite{gallier15
} point out the strong influence of confinement on the suspension rheology.\\
Several studies of dry granular media demonstrate that vibrations may induce volume fraction instabilities~\cite{krengel13} and the build-up of patterns has already been observed in oscillating suspensions, both experimentally \cite{gondret96, moosavi14} and in numerical simulations~\cite{loisel15}. These latter studies dealt with unbounded dilute suspensions (volume fraction $\phi \leq 10\%$) of buoyant beads ($\Delta \rho \neq 0$) oscillating at moderately high frequencies ($f \geq 6\, \mathrm{Hz}$).\\
In the present study, we analyze experimentally the behavior of a non-Brownian, iso-dense suspension of spheres submitted to a periodic low frequency square wave oscillation of the flow in a Hele-Shaw cell of gap $H$. We do observe an instability of the initially homogeneous suspension in the form of stripes of different particle concentrations transverse to the flow. We map the domain of existence of this instability in the space of the control parameters of the experiment, i.e. the cell thickness $H$, the amplitude $A$ and period $T$ of the fluid oscillations. We determine the variations of the wavelength, $\lambda$, and of the critical onset time, $t_c$, of this instability as a function of these control parameters.  Using  fluorescent dye dissolved in the fluid, we map the particle fraction across the gap $H$ and observe that the instability is associated to a migration of the particles from the center of the cell towards its walls. In order to account for the main features of this instability, we use the theory of longitudinal instability due to normal stresses difference \cite{brady02} and recent measurements of the dependence of the first normal stresses difference  on concentration\cite{dbouk13,gallier16}.

\section{Experimental set-up and data processing}

\begin{figure}[hptb]
\centering
\includegraphics[width=0.6\textwidth,clip]{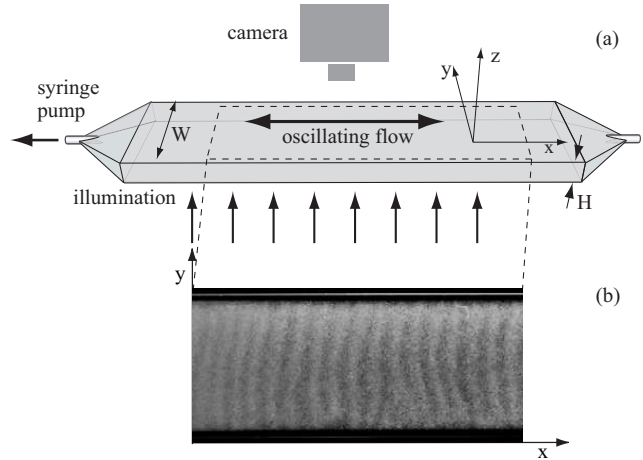}
\caption{ a) Schematic view of the experimental set-up.  b) Top view of a part of the instability pattern. Mean volume fraction $\phi=0.35$; cell gap $H=0.4 \,\mathrm{mm}$, the cell width, $W=8\,\mathrm{mm}$, corresponds to the distance between the two black boundaries.}
\label{insta-image}       
\end{figure}

The experimental set-up is shown on Fig.~\ref{insta-image}-a \cite{roht17}. The iso-dense suspension consists of polystyrene beads of diameter $d = 40\, \mu\mathrm{m}$ immersed in a water glycerol mixture ($21\%$ mass concentration of glycerol) of viscosity $\eta = 1.8 \, \mathrm{mPa\,. s}$ and density $\rho = 1050 \, \mathrm{kg\, m^{-3}}$ matching that of the particles at a temperature of $23^\circ \, \mathrm{C}$. The particles are almost mono-disperse with a root mean square diameter deviation of less than a few $\%$. For visualization across the gap \cite{Borrero16}, suspensions of
$40 \mathrm{\mu m}$ and $60 \mathrm{\mu m}$ \textit{PMMA} particles has been used. The suspension is contained in  Hele-Shaw cells of length $L = 100\, \mathrm{mm}$, and of different gap thicknesses $H$ ranging from $0.3 \, \mathrm{mm}$ to $1.2 \, \mathrm{mm}$; the aspect ratio is larger than $10$.

The oscillating flow of the suspension is induced by a computer controlled syringe pump fitted with $2.5\, \mathrm{ml}$ glass syringes. We use a symmetrical square wave variation of the flow rate. The period $T$  varied from $0.4\, \mathrm{s}$ to $10\, \mathrm{s}$; the lower value is set largely by the limited frequency response of the tubing and the syringe. The amplitude $A$ of the mean displacement of the fluid in the cell ranged from $A \simeq 0.5\, \mathrm {mm}$ to $A = 15\, \mathrm{mm}$ with a mean displacement $A$ in the first half of the period and $-A$ in the other half; the change of directions of the syringe took place in less than $0.1\,s$. A few experiments have been performed with a sine wave variation of the flow rate and a similar instability has been observed in some cases: it was however less clear-cut than for square waves, possibly due to the non constant velocity between flow reversals.

The cell is horizontal (thickness parallel to the vertical direction $z$) and illuminated from below. Images of the  patterns induced by the instability are acquired by a Nikon D800 camera located at 10 cm above the cell
and used in the movie mode: it captures $25$ frames per second with a $1920 \times 1080$ pixels resolution.
\begin{figure}[hptb]
\centering
\includegraphics[width=0.7\textwidth]{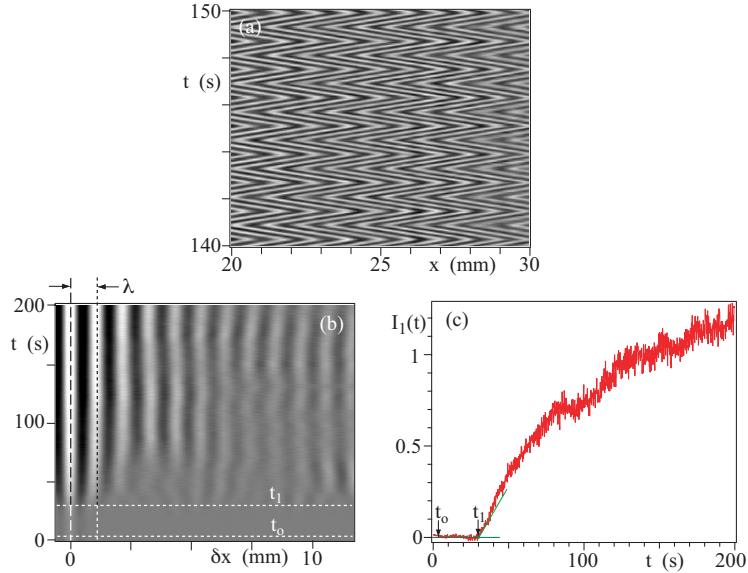}
\caption{a) Top:  Spatiotemporal diagram of the local transmitted light intensities averaged over an interval $\Delta y = 0.1 W$ in the transverse direction (grey level); vertical scale: time; horizontal scale: coordinate $x$ along the flow direction.  b) Spatiotemporal diagram of the autocorrelation function $F$ (grey levels) of the profiles of the above figure; vertical scale: time; horizontal scale: interval $\delta x$ (Eq.~\ref{autocorel}); the peaks are separated by one wavelength, $\lambda$. c) Plot of the intensity of the second peak of the autocorrelation function (at $\delta x = \lambda = 1.05\,\mathrm{mm}$) versus time; flow starts at time $t_0$; $t_1$ corresponds to the appearance of the instability;  $t_c = t_1-t_0$ characterizes the  time lapse for the onset of the instability. $\phi = 0.35$, $H=0.5\,\mathrm{mm}$, $T = 1.2\, \mathrm{s}$ and $A = 2.65 \, \mathrm{mm}$.}
\label{fig-streaks}       
\end{figure}

In the experimental procedure, the suspension obtained after mixing is injected into the Hele-Shaw cell; the uniformity of the transmitted light is a test of the constancy of the mean concentration (averaged over the cell thickness)  over the whole cell. The reproducibility of the experimental results requires a well defined protocol in order to start from the same initial conditions with an homogenous concentration, not only over the whole cell but also across its aperture. This homogeneity has been  tested by a visualization across the gap, using an index-matched suspension in a fluorescent fluid. The protocol, determined by trial and error, is to inject slowly the mixed suspension, wait a few minutes and then start the experiment. An important point is also to eliminate all traces of the pattern of the previous experiment: for this purpose, we induce slow, large scale oscillations of the fluid volume.  Then, with a suspension initially at rest, we start the square wave oscillations at the time $t_o$.
After a time lapse $t_c$, stripes transverse to the flow appear and reach a stationary shape and contrast within a few tens of seconds.
Fig.~\ref{insta-image}-b displays a typical top view of a part of the cell after this stationary regime has been achieved: the instability is marked by the appearance of periodic stripes transverse to the flow and of wavelength $\lambda$. These stripes correspond to a modulation of the transmitted light intensity due to variations of the particle concentration. As shown by Fig.~\ref{insta-image}b, the bands are more visible and straighter in the region of the axis of symmetry: in order to reduce the influence of the noise of the image, we average, for a given value of $x$, the grey levels of all pixels located within a range of distances of width  $\Delta y \sim 0.1 W$ centered on the axis. The visibility of the stripes on the variation curve of this average with $x$  is  enhanced by subtracting out the low frequency variation components  due, for instance, to the inhomogeneities of the illumination (these components are estimated by applying a smoothing filter to the original curve). The profiles $I(x,t)$ obtained in this way at different times $t$ are  plotted as grey levels in the spatiotemporal diagram of Fig.~\ref{fig-streaks}a. The zigzag structures visible in the diagram reflects the displacement of the bands along $x$ induced by the periodic flow; as expected, the segments of the structure have a linear shape due to the constant value of the flow rate between flow reversals. Finally, we compute, for all times $t$,  the autocorrelation  function:
 \begin{equation} \label{autocorel}
 F(\delta x, t) = \int I(x,t)\times I(x-\delta x, t)\, \mathrm{d}x.
 \end{equation}
 The function $F(\delta x, t)$ is plotted as grey levels
 in the spatiotemporal diagram of  Fig.~\ref{fig-streaks}b as a function of time and of the interval $\delta x$. $F(\delta x, t)$ is symmetrical with respect to $\delta x = 0$ so that only the right part of the diagram is plotted. The diagram displays alternate bright and dark vertical bands: their period corresponds to the wavelength $\lambda$ of the variations of the volume fraction induced by the instability. The visibility of the bands decreases with $\delta x$ over a distance which characterizes the spatial correlation  of the stripe pattern: depending on the experiment, it may range from $\sim 5 \lambda$ to $\sim 30 \lambda$. Practically, $\lambda$ is determined by plotting the distances $\delta x$ corresponding to the different maxima of $F$ as a function of their number in the sequence and performing a linear regression. With the protocol we have used,  the values of  $\lambda$ are reproducible to within  typically $\pm 10\%$ for a given set of control parameters. As seen on Fig.~\ref{fig-streaks}b,  the amplitude of the variations of the autocorrelation function increases with time from $t = t_0$ onwards. In Fig.~\ref{fig-streaks}c,  the intensity of the second peak of the autocorrelation function ($\delta x = \lambda$) is plotted versus time (flow starts at $t = t_0$). Time $t_1$ corresponds to the onset of instability because the appearance of this peak coincides with that of the periodic structure: as a result, $t_c = t_1-t_0$ measures the time needed for the instability to develop with an accuracy of a few seconds.
\section{Experimental results}
\label{results}
\begin{figure}[htbp]
\centering
\includegraphics[width=0.5\textwidth]{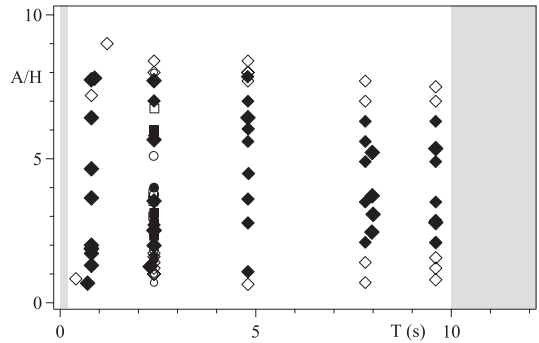}
\caption{Map of the range of observation of the stripe pattern as a function of the reduced amplitude $A/H$ of the mean fluid displacement and of the period $T$. Particle fraction $\phi = 0.35$. Different symbol shapes correspond to gap values: $H =  0.4 \,\mathrm{mm}$ $(\square)$,  $H = 0.5\, \mathrm{mm}$ $(\triangle)$, $H = 0.7 \,\mathrm{mm}$ $(\Diamond)$, $H = 1 \,\mathrm{mm}$ $(\bigcirc)$. Solid (open) symbols correspond respectively to experiments in which the instability is (is not) observed. Grey shades: domains of values of $T$ for which no experiments were performed.}
\label{DP}
\end{figure}
The control parameters  in these experiments are the particle volume fraction, $\phi$, the cell thickness, $H$, the period, $T$ and the peak to peak amplitude $A$ of the mean displacement of the fluid. The measurements are the wavelength $\lambda$ of the stripes pattern and the time $t_c$ corresponding to the onset of instability. Let us analyze step by step the influence of each control parameter.\\
{\it - Particles concentration $\phi$}\\
Instabilities were observed only for suspensions of large enough volume fraction  $\phi \geq \phi_{min}=0.25$ and up to $\phi = 0.35$. This latter upper limit of $\phi$
reflects the lack of confidence in the uniformity of the suspension at larger concentrations and the reduced intensity  of the transmitted light which makes the patterns less visible. In additional experiments, we observed that a non-zero density contrast between the beads and the fluid hinders the appearance of the instability rather than fostering it. For a relative density contrast $\Delta \rho/\rho \geq 3\%$, the periodic pattern does not appear any more. In all experiments discussed below, the density contrast is zero within $1\,\mathrm{^o\mkern-5mu/\mkern-3mu_{oo}}$ relative accuracy. \\
{\it - Ratio of cell thickness $H$ by sphere diameter $d$.} \\
For the suspension of $d=40 \, \mu \mathrm{m}$ diameter spheres, we observed the instability pattern for all cell thicknesses except for $H=0.3 \,  \mathrm{mm}$ ($H/d\sim 7.5$). To test this lower bound the value of $H/d$, we used the same type of spheres but with a larger diameter $d=60 \, \mathrm{\mu m}$: in this case, the instability was not observed in the $0.4\,   \mathrm{mm}$ cell and only for $H \geq 0.5 \,   \mathrm{mm}$. Therefore, a thick enough cell with $H/d \geq 8$ is required for the instability to occur: this might suggest a continuous coarse graining.\\
{\it - Amplitude $A$ and period $T$ of fluid displacements.}\\
The instability pattern has been observed in a broad range of $A$ and $T$ values for different cell gaps and concentrations satisfying the conditions discussed above. Note that the instability was still visible for the shortest period $T = 0.4\, \mathrm{s}$ compatible with our experimental set up: we cannot therefore determine whether there is a lower limit to the period for observing the instability. The diagram of existence is displayed in Fig.~\ref{DP} with  $A/H$ and $T$ as vertical and horizontal coordinates. The data points correspond to a volume fraction $\phi = 35\%$ and to different cell gaps: $0.4\, \mathrm{mm} \leq H \leq 1 \, \mathrm{mm}$. As shown in this figure,
for all periods used ($T \leq 10 \, s $), the instability only occurred  in a finite range of amplitudes ($A_c \leq A \leq A_l$) with both an upper limit and a non zero lower one. The domain of existence of the instability seems to become slightly narrower as $T$ increases from $1$ to $10\, s$.\\
{\it - Dependence of the wavelength $\lambda$ on $T$ and $A$.}\\
\begin{figure}[htbp]
\centering
\includegraphics[width=1\textwidth]{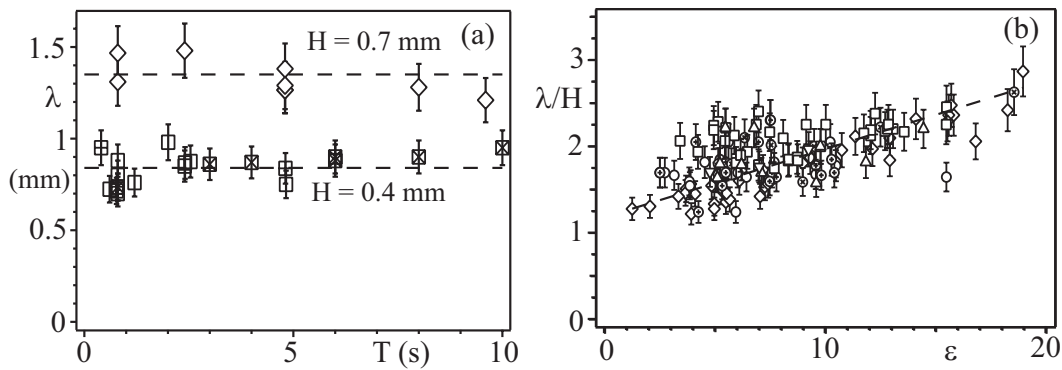}\\
\caption{a) Wavelength $\lambda$ of the instability pattern versus the period $T$  for two different cell gaps: $H =0.4 \, \mathrm{mm}$  and  $H=0.7\, \mathrm{mm}$.
Symbol fillings indicate particles volume fractions: $\phi = 25\%\, (+); \,30\%\, (\centerdot),\, 35\%\, \mathrm{(open\, symbols)}$. b) Dimensionless wavelength $\lambda /H$ versus  strain deformation $\epsilon= A/(H/2)$ for different cell gaps: $0.4\, \mathrm{mm}\leq H\leq1.2 \, \mathrm{mm}$ and volume fractions: $0.25\leq \phi \leq 0.35$. Symbol fillings: $40\, \mathrm{\mu m}$ polystyrene particles (open symbols), $40\, \mathrm{\mu m}$ \textit{PMMA} particles $(+)$, $60\, \mathrm{\mu m}$ \textit{PMMA} particles $(\times)$.
In both graphs (a) and (b),  the different symbol shapes correspond to the same $H$ values as in Fig.~\ref{DP} with, in addition,  $(\triangledown)$  for $H = 1.2\, \mathrm{mm}$. Error bars reflect the typical dispersion of the experimental results ($\pm 10\,\%$) for experiments using a same set of control parameters.}
\label{lambda}       
\end{figure}

Fig.~\ref{lambda}a displays the variation of the wavelength $\lambda$ of the instability pattern with the period $T$ of the flow for two different gap thicknesses $H=0.4 \,\mathrm{mm}$  and $H=0.7\,\mathrm{mm}$ and for similar ranges of strain deformations $\epsilon$ in both cases (respectively $9\lesssim  \epsilon \lesssim 12.5$ and $6.5 \lesssim \epsilon \lesssim 12.5$); here, and in the following, the mean strain deformation during one half period is taken equal to $\epsilon = 2A/H$ in order to take into account the symmetry of the Poiseuille profile. The wavelength is  independent of  $T$ and the ratio $\lambda/H\simeq 2 \pm 0.2$ is similar for both thicknesses. In view of these results, we have plotted in Fig.\ref{lambda}b  for different $H$ and $\phi$ the dimensionless wavelength, $\lambda/H$ as a function of $\epsilon$. Comparing Figs. \ref{lambda}a and \ref{lambda}b shows that using the dimensionless variable $\lambda/H$ instead of $\lambda$ improves the collapse of the data corresponding to different gaps $H$  and validates the use of $H$ as the scaling variable for $\lambda$. This collapse is retained at all values of $A$ when the deformation $\epsilon$ is used as the horizontal scale: the slow increase of  $\lambda$ with $\epsilon$ may be fitted by $\lambda/ H \simeq 1.2 +0.08 \, \epsilon$ (dashed line in Fig.~\ref{lambda}).\\
{\it -Characteristic time lapse $t_c$ for the appearance of the instability}\\
\begin{figure}[htbp]
\centering
\includegraphics[width=0.5\textwidth]{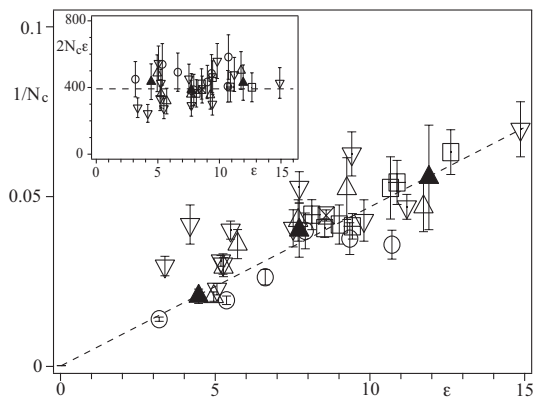}\hskip24pt
\caption{Main graph: evolution of $1/N_c=T/t_c$ (inverse of the number of flow periods before the appearance of the instability pattern) with the  deformation $\epsilon$
during one half period. Dashed line:  linear fit with the data (see text). Insert : variation with $\epsilon$ of  the total accumulated deformation $2\,N_c \,\epsilon$ at the onset of the instability. Horizontal dashed line: linear fit to the data (see text). In both graphs, symbol shapes have the same meaning as in Figs.~\ref{DP}  and \ref{lambda}. Symbol fillings correspond to the periods:  $T = 0.8\, \mathrm{s}$ (solid symbols),  $1.2\, \mathrm{s}$ (open symbols), $1.8 \, \mathrm{s}$ $(+)$, $2.4 \, \mathrm{s}$ $(\centerdot)$,  $4.8\, \mathrm{s}$ $(\times)$.}
\label{seuil}       
\end{figure}

In the experiments for which the instability has been observed, we measured the characteristic time lapse $t_c$ before its appearance (see Fig.~\ref{fig-streaks}c) for different $H$, $T$ and $A$ values. In order to characterize the onset of the instability by a dimensionless expression, we have studied the variations of the ratio $T/t_c$ (instead of $t_c$) representing the inverse of the number $N_c$ of oscillations above which the instability becomes visible. In Fig. \ref{seuil}, we plot $1/N_c$ versus the characteristic deformation $\epsilon$ defined above.
We observe that $1/Nc$ follows a roughly linear increasing trend as a function of $\epsilon$ with a similar proportionality coefficient for all values of $H$ and $T$ used  here  so that $1/N_c \simeq 0.005 \,\epsilon$. This means that the minimum total deformation of the sheared suspension accumulated over $N_c$ periods for observing the instability is $ 2\,N_c\,\epsilon \simeq 400$ (see insert). If, for further use, we assume that $1/t_c$ scales like the instability growth rate $\sigma$, we obtain: $\sigma \simeq 1/t_c\simeq 0.0025 \, \dot{\gamma}$ where $\dot{\gamma}= 2 \epsilon /T$ is an effective shear rate.
\section{Observations across the gap: particle migration and nature of the instability}
\begin{figure}[htbp]
\centering \includegraphics[width=\textwidth]{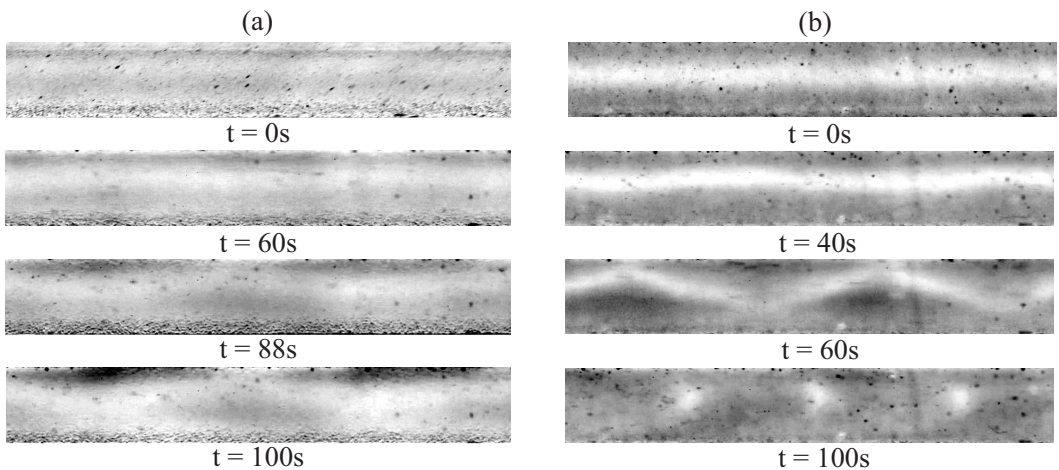}
\caption{Development of the instability across the gap of a cell with $H=1.2 \,\mathrm{mm}$ illuminated by a thin plane  light sheet ($T = 1.2\,\mathrm{s}$, $A = 5\,\mathrm{mm}$). All images are snapshots of the fluorescence intensity in the cell gap: brighter zones correspond to a lower particle concentration. Graph (a): development of the instability with the time $t$ elapsed after the pump is set in motion with a weak segregation in the initial state. Graph (b): development of the instability with a strong initial depletion of the particles in the center of the gap (bright band in the middle) and an  initial accumulation near the walls.}
\label{gap}       
\end{figure}

Using the setup of Fig.\ref{insta-image}, we have analyzed the global features of the  instability, like its diagram of existence and the dependence of its wavelength on the experimental control parameters. However, in order to understand the physical process(es) at work, more local experimental observations and measurements are needed. For that purpose, we  study now the time dependence of the distribution of the particle concentration across the gap of the  cell during the development of the instability. In order to observe this concentration variation, we use both a \textit{PMMA} cell and \textit{PMMA} particles of diameter $40 \, \mathrm{\mu m}$ and concentration: $35\,\%$ immersed in a fluid of the same density and refractive index as \textit{PMMA}\cite{Borrero16}. We dissolve a fluorescent dye (rhodamine) in the fluid and illuminate the cell with a green laser sheet ($532\, \mathrm{nm}$) perpendicular to the cell walls and parallel to the  plane $(x,y)$ in Fig.~\ref{insta-image}. The camera axis is  parallel to the axis $y$ so that it maps the particle concentration variations across the gap inside the illuminated plane (a higher fluorescence intensity corresponds  to a smaller concentration of the non fluorescent particles).

Both Figs.~\ref{gap}a and \ref{gap}b  display a series of snapshots showing the development of the instability with time starting from the initial distribution when the pump is started (upper pictures); Fig.~\ref{gap}a corresponds to an initial distribution obtained using our standard experimental procedure and do not displays a clear segregation of the particles in the gap (the light intensity is similar near the walls and in the center of the gap). In the series of
Fig.~\ref{gap}b, instead, there is in the initial state (top image) a strong depletion of the particles in the middle of the gap (straight bright band) and a strong accumulation near the walls (dark regions). The relative maximum difference between the intensities of fluorescence in the bright and dark zones is found to be of the order of $15\%$. Assuming a roughly linear relation between the light intensity and the fraction of liquid and multiplying by the mean liquid volume fraction $1-\phi = 0.65$, this provides a rough estimation $\Delta \phi \sim 10\%$ of the concentration contrast between the vicinity of the walls and the center of the gap in this particular experiment.\\
In the first series (a), the second picture ($t = 60\,\mathrm{s}$) displays a band of slightly lower concentration in the gap. This band is more visible and oscillates at $t = 88\,\mathrm{s}$ while  accumulation zones appear near the walls. Both the waviness of the band of lower concentration and the particle accumulation in regions close to the walls are strongly amplified at $t = 100\,\mathrm{s}$. It is difficult to determine the relative contributions of the development of the instability and of the segregation process.\\
In the second series of pictures (Fig.~\ref{gap}b), there is instead a strong initial segregation with a depletion of particles in a narrow stripe (width $\sim H/4$) in the middle of the gap.  This stripe displays already waviness at $t = 40 \, \mathrm{s}$ which grows to a large amplitude at $t = 60 \, \mathrm{s}$  and develops finally into a cellular structure ($t = 100 \, \mathrm{s}$).\\
These two observations of the distribution of the particle concentration across the gap  suggest therefore  an instability scenario of particles migration towards the wall followed by (or associated with) an instability due to particle concentration gradients with a higher concentration at the walls.
\section{Discussion and interpretation}
From the present results, particle migration from the middle of the gap to the walls is likely to be the key condition for observing the stripes instability. Migration of particles in flowing suspensions has been observed for a long time \cite{parsi87, leighton87, leighton87b}. In continuous pressure driven flows in a pipe, the migration is from the walls towards the center \cite{ lyon98} or, more precisely, from the high shear regions close to the walls to low shear region towards the pipe center line. This effect was modeled using the so-called shear induced migration model \cite{leighton87,phillips92} or by introducing normal stresses within the flowing suspension \cite{nott94,morris99}. When a suspension is oscillating in a pipe, "anomalous" migration of particles from the center to the walls was predicted \cite{morris01} and confirmed experimentally \cite{butler99}. For large strains, the "normal" migration to the center is recovered. In our experiments we also observed this anomalous migration (leading to the stripes instability) for deformations smaller than $\epsilon \leq 15$ .
These instabilities are however only observed up to a maximum deformation (Fig. \ref{DP}) which may reflect the approach of a change of migration mode; this is consistent with "normal" migration from the walls towards the center observed for continuous flows.
In our experiments, the instability is observed for strains $\epsilon \leq 15-20$ much larger  than those  ($\epsilon \leq 0.1-0.2$) used in previous numerical \cite{morris01} and experimental \cite{butler99} works (in this latter case  in a cylindrical geometry). \\
The remaining question to address is the mechanism responsible for the instability of this anomalous spatial distribution leading to the stripes pattern (Fig.~\ref{gap}b). A similar question was addressed for the instability of the interface between non-Newtonian fluids by Brady and Carpen~\cite{brady02}. At the interface between the two fluids, owing to their rheological properties, the normal stresses perpendicular to the interface need not be continuous; as a result, if the initially flat interface is perturbed, the mismatch of the normal stresses creates a traction that can destabilize the interface. The second normal stress difference ($N_2=\sigma_{zz}-\sigma_{yy}$) governs  the instability components transverse to the flow while the first normal stresses difference ($N_1=\sigma_{xx}-\sigma_{zz}$) governs perturbations in the flow direction; the latter issues were also analyzed for elastic fluids \cite{hinch92}. A necessary condition \cite{brady02,hinch92} for observing an instability in the flow direction is to have a larger $N_{1}$ close to the wall than in the middle, namely $N_1 (z=0)=N_1 (z=H)>N_1(z=H/2)$. However, for concentrated suspensions, the magnitude and the sign of $N_1$ are still under debate \cite{brady02,gallier16,dbouk13}. Recent experiments \cite{dbouk13} and numerical simulations \cite{gallier16} in dense suspensions predict a positive value of $N_1$ increasing with the particle volume fraction ($\partial N_{1}/ \partial \phi > 0$) and almost proportional to the shear rate ($\dot{\gamma}$). Note that, in a Poiseuille flow, the shear rate is larger at the wall than in the middle of the gap. In the experiments for which the instability was observed across the gap (Fig.~\ref{gap}b), the particle concentration is larger at the wall than in the middle of the cell ($\phi (z = 0\, \mathrm{or}\, H) > \phi (z=H/2)$). Hence, for a positive $N_1$ increasing with both the shear rate and the concentration, one has $N_1 (z = 0\, \mathrm{or}\, H) > N_1 (z=H/2)$ which promotes the instability in the flow direction \cite{brady02,hinch92}. The experimental variation of the growth rate with the shear rate, $\sigma \simeq 1/t_c\simeq 0.0025 \, \dot{\gamma}$ is coherent with the expected theoretical prediction \cite{brady02}.
\section{Conclusion}
We have analyzed experimentally the behavior of a non-Brownian, iso-dense suspension of spheres submitted to periodic square wave oscillations of the flow rate (period $T$ and amplitude of the fluid displacement $A$) in a Hele-Shaw cell of gap $H$. We do observe an instability of the initially homogeneous concentration in form of concentration variation stripes transverse to the flow. The wavelength of these regular spatial structures scales roughly as the gap of the cell and is independent of the particle concentration and of the period of oscillation. This instability requires large enough particle volume fractions $\phi \geq 0.25$ and a gap large enough compared to the spheres diameter ($H/d \geq 8$); it is observed in a rather broad range of periods of the square wave and for amplitudes of the fluid displacement between lower and upper bounds. The analysis of the particle concentration distribution across the
gap supports a two steps scenario of the instability: migration of the particles towards the cell walls followed by an instability due to the concentration gradient across the gap. Recent measurements of the first normal stresses difference \cite{dbouk13,gallier16} and the theory of longitudinal instability due to the first normal stresses difference \cite{brady02} account reasonably for our experimental observations.
More quantitative comparisons will require further measurements of the concentration across the gap as well as of the particle distribution function in order to estimate $N_1(z)$. It will also be important to investigate further the influence of small density contrasts on the instability. This instability might be described  in the spirit of what is done for miscible fluid interfaces \cite{talon11} by introducing a pseudo-interface.
\acknowledgments
We acknowledge support by the LabEx PALM (ANR-10-LABX-0039-PALM),
by the LIA PMF-FMF (Franco-Argentinian International Associated Laboratory
in the Physics and Mechanics of Fluids) and from
 UBACyT 20020130100570BA. The thesis of Y.L. Roht is supported by a fellowship
 from the Peruilh foundation of the Faculty of Engineering of the Buenos-Aires University, by
 a Bec.Ar fellowship  and by an Eiffel fellowship.
The authors want to thank A. Aubertin, L. Auffray and R. Pidoux for their help in the design
and realization of the experimental set-up.
\bibliographystyle{eplbib}


\begin{thebibliography}{10}
\expandafter\ifx\csname url\endcsname\relax\def\url#1{\texttt{#1}}\fi

\bibitem{leighton87}
\Name{Leighton D. \and Acrivos A.} \REVIEW{Journal of Fluid
  Mechanics}{181}{1987}{415}.

\bibitem{phillips92}
\Name{Phillips R.~J., Armstrong R.~C., Brown R.~A., Graham A.~L. \and Abbott
  J.~R.} \REVIEW{Physics of Fluids A: Fluid Dynamics}{4}{1992}{30}.

\bibitem{nott94}
\Name{Nott P.~R. \and Brady J.~F.} \REVIEW{Journal of Fluid
  Mechanics}{275}{1994}{157}.



\bibitem{butler99}
\Name{Butler J.~E., Majors P.~D. \and Bonnecaze R.~T.} \REVIEW{Physics of
  Fluids}{11}{1999}{2865}.

\bibitem{hinch92}
\Name{Hinch E., Harris O. \and Rallison J.} \REVIEW{Journal of Non Newtonian
  Fluid Mechanics}{43}{1992}{311}.

\bibitem{brady02}
\Name{Brady J.~F. \and Carpen I.~C.} \REVIEW{Journal of Non-Newtonian Fluid
  Mechanics}{102}{2002}{219}.

\bibitem{snook16}
\Name{Snook B., Butler J.~E. \and Guazzelli {\'E}.} \REVIEW{Journal of Fluid
  Mechanics}{786}{2016}{128}.

\bibitem{gallier15}
\Name{Gallier S., Lemaire E., Peters F. \and Lobry L.} \REVIEW{Phys. Rev.
  E}{92}{2015}{020301}.

\bibitem{krengel13}
\Name{Krengel D., Strobl S., Sack A., Heckel M. \and P{\"o}schel T.}
  \REVIEW{Granular Matter}{15}{2013}{377}.

\bibitem{gondret96}
\Name{Gondret P. \and Petit L.} \REVIEW{Physics of Fluids}{8}{1996}{2284}.

\bibitem{moosavi14}
\Name{Moosavi R., Maleki M., Shaebani M.~R., Ruiz-Su{\'a}rez J.~C. \and
  Cl{\'e}ment E.} \REVIEW{EPL (Europhysics Letters)}{107}{2014}{34006}.

\bibitem{loisel15}
\Name{Loisel V., Abbas M., Masbernat O. \and Climent E.} \REVIEW{Physics of
  Fluids}{27}{2015}{}.

 \bibitem{dbouk13}
\Name{Dbouk T., Lobry L. \and Lemaire E.} \REVIEW{Journal of Fluid
  Mechanics}{715}{2013}{239}.

 \bibitem{gallier16}
\Name{Gallier S., Lemaire E., Lobry L. \and Peters F.} \REVIEW{Journal of Fluid
  Mechanics}{799}{2016}{100}.

\bibitem{roht17}
\Name{Roht Y., Gauthier G., Hulin J., Salin D., Chertcoff R., Auradou H. \and
  Ippolito I.}  in \Book{Powders and Grains 2017 -- 8th International
  Conference on Micromechanics on Granular Media}, edited by \Name{{EPJ Web of
  Conferences}} \textbf{140} (2017) 09029.

\bibitem{Borrero16}
\Name{Borrero-Echeverry D. \and Morrison B. C.~A.} \REVIEW{Experiments in
  Fluids}{57}{2016}{123}.

\bibitem{parsi87}
\Name{Parsi F. \and Gadala-Maria F.} \REVIEW{J. Rheol.}{31}{1987}{725}.

\bibitem{leighton87b}
\Name{Leighton D. \and Acrivos A.} \REVIEW{Journal of Fluid
  Mechanics}{177}{1987}{109}.

\bibitem{lyon98}
\Name{Lyon M.~K. \and Leal L.~G.} \REVIEW{Journal of Fluid
  Mechanics}{363}{1998}{25}.

\bibitem{morris99}
\Name{Morris J. F. \ and Boulay F.} \REVIEW{J. Rheol.}{43}{1999}{1213}.

\bibitem{morris01}
\Name{Morris J.~F.} \REVIEW{Physics of Fluids}{13}{2001}{2457}.


\bibitem{talon11}
\Name{Talon L. \and Meiburg E.} \REVIEW{J. Fluid Mech.}{686}{2011}{484}.

\end{thebibliography}

\end{document}